\documentclass[journal,twoside,web]{ieeecolor}
\usepackage{generic}
\usepackage{amsmath,amssymb,amsfonts}
\usepackage{algorithmic}
\usepackage{graphicx}
\usepackage{textcomp}

\usepackage{url}
\usepackage[hidelinks]{hyperref}
\usepackage{placeins}
\usepackage[T1]{fontenc}
\usepackage{hyperxmp}
\usepackage{booktabs}
\usepackage{balance}
 \usepackage{array,multirow}

\newcommand{\isep}{\mathrel{{.}\,{.}}\nobreak}
\newcommand{\sig}{\textsuperscript{$\dagger$}\hspace{0.0ex}}

\hypersetup{
	pdfcopyright={
		This work uses FontAwesome Icons by Fonticons, Inc. licensed under a CC-BY 4.0 License https://creativecommons.org/licenses/by/4.0/ and FontAwesome fonts licensed under a SIL OFL 1.1 License (see below). Copyright (c) 2019, Fonticons, Inc., https://fontawesome.com This Font Software is licensed under the SIL Open Font License, Version 1.1. This license is copied below, and is also available with a FAQ at: https://scripts.sil.org/OFL SIL OPEN FONT LICENSE Version 1.1 - 26 February 2007 PREAMBLE The goals of the Open Font License (OFL) are to stimulate worldwide development of collaborative font projects, to support the font creation efforts of academic and linguistic communities, and to provide a free and open framework in which fonts may be shared and improved in partnership with others. The OFL allows the licensed fonts to be used, studied, modified and redistributed freely as long as they are not sold by themselves. The fonts, including any derivative works, can be bundled, embedded,  redistributed and/or sold with any software provided that any reserved names are not used by derivative works. The fonts and derivatives, however, cannot be released under any other type of license. The requirement for fonts to remain under this license does not apply to any document created using the fonts or their derivatives. DEFINITIONS "Font Software" refers to the set of files released by the Copyright Holder(s) under this license and clearly marked as such. This may include source files, build scripts and documentation. "Reserved Font Name" refers to any names specified as such after the copyright statement(s). "Original Version" refers to the collection of Font Software components as distributed by the Copyright Holder(s). "Modified Version" refers to any derivative made by adding to, deleting, or substituting — in part or in whole — any of the components of the Original Version, by changing formats or by porting the Font Software to a new environment. "Author" refers to any designer, engineer, programmer, technical writer or other person who contributed to the Font Software. PERMISSION \& CONDITIONS Permission is hereby granted, free of charge, to any person obtaining a copy of the Font Software, to use, study, copy, merge, embed, modify, redistribute, and sell modified and unmodified copies of the Font Software, subject to the following conditions: 1) Neither the Font Software nor any of its individual components, in Original or Modified Versions, may be sold by itself. 2) Original or Modified Versions of the Font Software may be bundled, redistributed and/or sold with any software, provided that each copy contains the above copyright notice and this license. These can be included either as stand-alone text files, human-readable headers or in the appropriate machine-readable metadata fields within text or binary files as long as those fields can be easily viewed by the user. 3) No Modified Version of the Font Software may use the Reserved Font Name(s) unless explicit written permission is granted by the corresponding Copyright Holder. This restriction only applies to the primary font name as presented to the users. 4) The name(s) of the Copyright Holder(s) or the Author(s) of the Font Software shall not be used to promote, endorse or advertise any Modified Version, except to acknowledge the contribution(s) of the Copyright Holder(s) and the Author(s) or with their explicit written permission. 5) The Font Software, modified or unmodified, in part or in whole, must be distributed entirely under this license, and must not be distributed under any other license. The requirement for fonts to remain under this license does not apply to any document created using the Font Software. TERMINATION This license becomes null and void if any of the above conditions are not met.  DISCLAIMER THE FONT SOFTWARE IS PROVIDED "AS IS", WITHOUT WARRANTY OF ANY KIND, EXPRESS OR IMPLIED, INCLUDING BUT NOT LIMITED TO ANY WARRANTIES OF MERCHANTABILITY, FITNESS FOR A PARTICULAR PURPOSE AND NONINFRINGEMENT OF COPYRIGHT, PATENT, TRADEMARK, OR OTHER RIGHT. IN NO EVENT SHALL THE COPYRIGHT HOLDER BE LIABLE FOR ANY CLAIM, DAMAGES OR OTHER LIABILITY, INCLUDING ANY GENERAL, SPECIAL, INDIRECT, INCIDENTAL, OR CONSEQUENTIAL DAMAGES, WHETHER IN AN ACTION OF CONTRACT, TORT OR OTHERWISE, ARISING FROM, OUT OF THE USE OR INABILITY TO USE THE FONT SOFTWARE OR FROM OTHER DEALINGS IN THE FONT SOFTWARE.
}
}

\def\BibTeX{{\rm B\kern-.05em{\sc i\kern-.025em b}\kern-.08em
    T\kern-.1667em\lower.7ex\hbox{E}\kern-.125emX}}
\begin{document}
\title{A Deep Learning Approach to Diagnosing Multiple Sclerosis from Smartphone Data}

\author{Patrick Schwab,  and Walter Karlen, \IEEEmembership{Senior Member, IEEE}
\thanks{This work was partially funded by the Swiss National Science Foundation (SNSF) project No. 167302 within NRP 75 ``Big Data''. We gratefully acknowledge the support of NVIDIA Corporation with the donation of the Titan Xp GPUs used for this research.  }
\thanks{P. Schwab and W. Karlen are with the Mobile Health Systems Lab, Institute of Robotics and Intelligent Systems, Department of Health Sciences and Technology, ETH Zurich, Switzerland (e-mails: patrick.schwab@hest.ethz.ch, walter.karlen@ieee.org). }
}

\maketitle

\begin{abstract}
Multiple sclerosis (MS) affects the central nervous system with a wide range of symptoms. MS can, for example, cause pain, changes in mood and fatigue, and may impair a person's movement, speech and visual functions. Diagnosis of MS typically involves a combination of complex clinical assessments and tests to rule out other diseases with similar symptoms. New technologies, such as smartphone monitoring in free-living conditions, could potentially aid in objectively assessing the symptoms of MS by quantifying symptom presence and intensity over long periods of time. Here, we present a deep-learning approach to diagnosing MS from smartphone-derived digital biomarkers that uses a novel combination of a multilayer perceptron with neural soft attention to improve learning of patterns in long-term smartphone monitoring data. Using data from a cohort of 774 participants, we demonstrate that our deep-learning models are able to distinguish between people with and without MS with an area under the receiver operating characteristic curve of {0.88} (95\% CI: 0.70, 0.88). Our experimental results indicate that digital biomarkers derived from smartphone data could in the future be used as additional diagnostic criteria for MS.
\end{abstract} 


\section{Introduction}
\label{sec:introduction}
\IEEEPARstart{M}{ultiple} sclerosis (MS) is a neurological disease that affects around 2 million people worldwide \cite{vos2016global}. The neural lesions caused by MS reduce the capability of neurons to transmit information, which leads to a wide range of symptoms, such as changes in sensation, mobility, balance, vision, and cognition \cite{brownlee2017diagnosis}. Diagnosing MS requires objective evidence of two lesions in the central nervous system disseminated both in time and space \cite{mcdonald2001recommended,polman2011diagnostic}. Physicians typically use a combination of clinical assessments of symptoms, blood tests, imaging, cerebrospinal fluid analysis and analysis of evoked potentials to rule out other diseases with similar symptoms \cite{mcdonald2001recommended,polman2011diagnostic,filippi2016mri}. Currently, no cure exists for MS, but there are treatments available that are effective at managing the symptoms of MS and may significantly improve long-term outcomes \cite{brownlee2017diagnosis,marziniak2016variations,scolding2015association}. To receive early access to these treatments, a timely diagnosis is of paramount importance for patients.

\begin{figure}[t]
\centerline{\includegraphics[width=\linewidth]{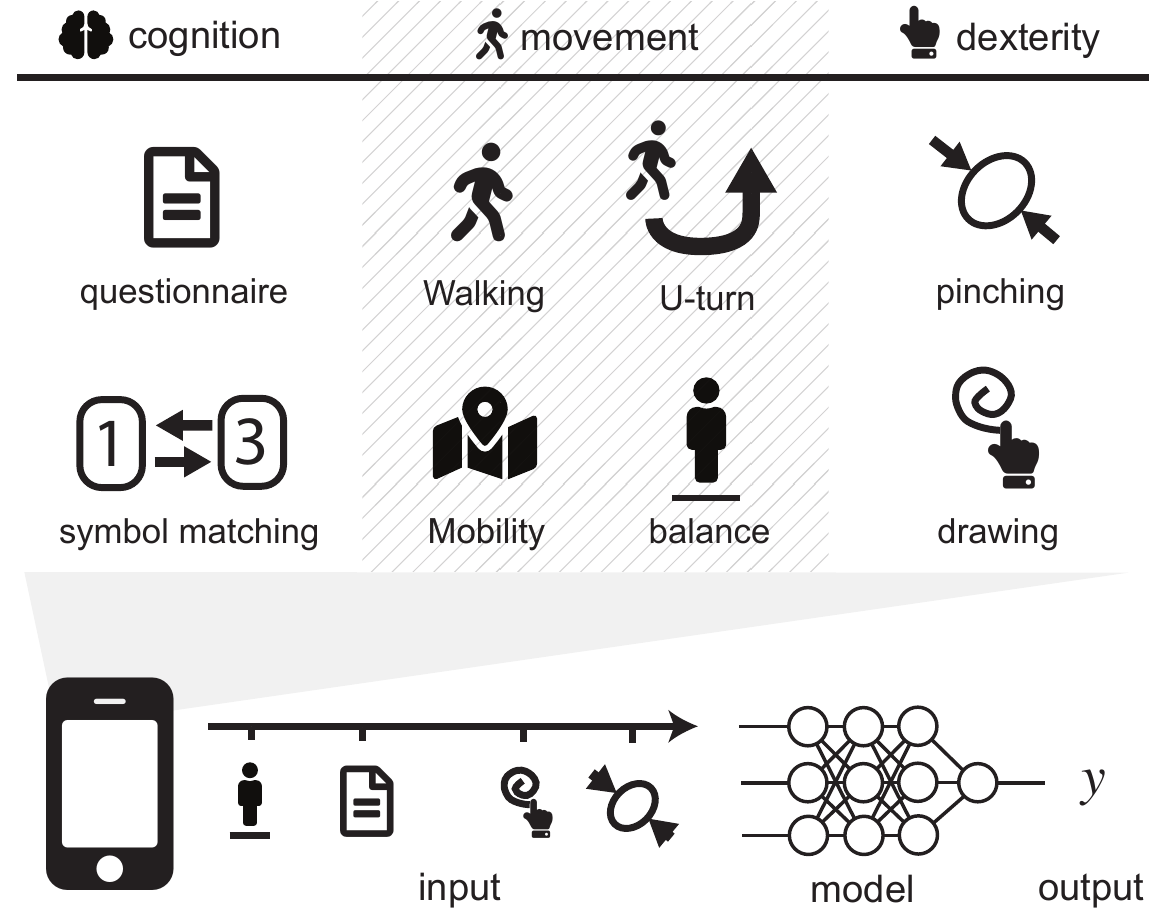}}
\caption{Smartphone-based tests (top) can be used to assess cognition, movement and finger dexterity symptoms of multiple sclerosis (MS) and track their progression over time. We train machine-learning models to learn to produce a scalar diagnostic score $y$ (bottom) from the data collected during any number of those tests to learn to diagnose MS. } 
\label{fig:teaser}
\end{figure}

Smartphone-based tests could potentially be used to quantify symptoms of MS in the wild over long periods of time (Figure \ref{fig:teaser}). However, to date, it has not been established whether and to what degree smartphone monitoring data can be used to derive digital biomarkers for the diagnosis of MS. A particularly challenging aspect of using smartphone data to derive digital biomarkers for the diagnosis of MS is that smartphone monitoring yields large amounts of high-resolution data from multiple symptom categories. Identifying the salient input segments and reaching a clinically meaningful conclusion from raw sensor data in an accurate and timely manner is therefore challenging both for physicians and machines.

To address these issues, we present a machine-learning approach for distinguishing between people with and without MS from smartphone data. At its core, our method uses an attentive aggregation model (AAM) that integrates the results of multiple tests over time to produce a single diagnostic score. By integrating neural attention in our model, we are additionally able to quantify the importance of individual tests towards the model's output. Our experiments on real-world smartphone monitoring data show that our method outperforms several strong baselines, identifies meaningful patterns, and that smartphone data could potentially be used to derive digital biomarkers for the diagnosis of MS.

Concretely, our contributions are as follows:
\begin{itemize}
\item We present a deep-learning approach to distinguishing between people with and without MS that integrates data from multiple types of smartphone-based tests performed over long time frames (up to more than 200 days).
\item We extend our deep-learning approach with neural soft attention in order to quantify the importance of individual input features towards the final diagnostic score.
\item We utilise real-world smartphone monitoring data from 774 subjects to evaluate, for the first time, the comparative performance of machine-learning models and several strong baselines in distinguishing between people with and without MS.
\end{itemize}

\section{Related Work}
Using machine learning to aid in medical tasks has attracted much research interest. Researchers have, for example, used machine learning for mortality modelling \cite{ghassemi2014unfolding}, sepsis decision support \cite{horng2017creating}, alarm reduction in critical care \cite{schwab2018not}, to provide explainable decisions in medical decision-support systems \cite{lundberg2017unified,lundberg2018explainable,schwab2018granger}, and to identify subtypes  in autism spectrum disorders \cite{doshi2014comorbidity} and scleroderma \cite{schulam2015clustering}. Giving a reliable diagnosis is one of the most challenging tasks in medicine that requires significant domain knowledge in the disease being assessed, and the ability to integrate information from a large number of potentially noisy data sources. Machine learning is an attractive tool to perform automated diagnoses because it can draw upon the experience of millions of historical samples in its training data, and seamlessly integrate data from multiple disparate data sources. In previous studies, machine learning has, for example, been used to diagnose skin cancer from image data \cite{esteva2017dermatologist}, glaucoma from standard automated perimetry \cite{chan2002comparison}, and a large number of diseases from electronic health records \cite{lipton2015learning,choi2016doctor} and lab test results \cite{razavian2016multi}. However, obtaining objective data about symptoms and symptom progression over time is challenging in many diseases. For some diseases, wearable devices and smartphones have emerged as viable tools for gathering diagnostic data in the wild. Smartphones and telemonitoring have, for example, been used to gather digital biomarkers for the diagnosis of melanomas \cite{wolf2013diagnostic}, bipolar disorder \cite{faurholt2019objective}, cognitive function \cite{piau2019current}, and Parkinson's disease \cite{tsanas2010enhanced,arora2015detecting,tsanas2010accurate,zhan2018using,schwab2019phonemd}. The use of machine learning on high-resolution health data often requires specialised approaches to effectively deal with missingness \cite{lipton2016directly,che2018recurrent}, long-term temporal dependencies \cite{choi2016doctor}, noise \cite{schwab2017beat}, heterogeneity \cite{libbrecht2015machine}, irregular sampling \cite{lipton2015learning}, sparsity \cite{lasko2013computational}, and multivariate input data \cite{schwab2018not,schwab2019phonemd,ghassemi2015multivariate}. In this work, we build on these advances to develop a novel approach to learning to diagnose MS from smartphone-derived digital biomarkers that addresses the aforementioned challenges.

\subsection{Monitoring and Diagnosis of MS} The clinical state-of-the-art in monitoring symptoms and symptom progression in MS is based on a combination of clinical assessments, such as neurological exams, magnetic resonance imaging (MRI), and the Expanded Disability Status Scale (EDSS) \cite{kurtzke1983rating,wattjes2015evidence}. However, these tests can only be performed at clinical centers by medical specialists. With dozens of mHealth apps available to manage MS on all major smartphone platforms, smartphone apps have recently emerged as a readily accessible alternative to non-invasively track symptoms of MS in the wild \cite{giunti2018supply,boukhvalova2019smartphone}. Prior studies on the use mHealth in MS have, for example, evaluated telemedicine-enabled remote EDSS scoring \cite{bove2018toward}, measurement devices for estimating walking ability \cite{dalla2017smart} and fatigue \cite{barrios2018msfatigue}, and machine learning for assessing gait impairment in MS \cite{mcginnis2017machine}. Epidemiologically, demographic factors, such as age and sex, have been shown to be predictive of MS \cite{pugliatti2006epidemiology}.

In contrast to existing works that focused on single daily-life aspects of already diagnosed MS patients, we present an approach to diagnosing MS from smartphone-derived digital biomarkers, and verify this approach on a real-world dataset collected from a MS cohort. Our machine-learning approach addresses multiple challenges in learning from sensor-based smartphone tests that are self-administered multiple times over long periods of time. Most notably, with the integration of a global neural soft attention mechanism, we enable the quantification of the importance of individual smartphone tests towards the final diagnostic score, overcome the challenges of missingness, sparse data, long-term temporal dependencies between tests, and multivariate data with irregular sampling. 

\section{Methods and Materials}
\subsection{Smartphone Tests} We utilise data collected by the Floodlight Open study, a large smartphone-based observational study for MS \cite{montalban2018floodlight,midaglia2019floodlight}. The de-identified dataset used in this work is openly available to researchers\footnote{\url{https://floodlightopen.com}}. In the study, participants were asked to actively perform a number of smartphone-based tests on a daily basis using their personal smartphones in the wild and without any clinical supervision (Figure \ref{fig:teaser}). However, participants were free to choose when and if they performed the daily tests. Many participants therefore did not strictly adhere to the daily test protocol, and performed the tests irregularly. In addition to the manual tests, the app also passively collected movement data of the participants in order to determine their radius of living. The following tests were included in the study (Figure \ref{fig:teaser}) \cite{montalban2018floodlight}: 
\begin{itemize}
\item[\raisebox{-.15\height}{\includegraphics[height=10pt]{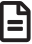}}]\textbf{Mood Questionnaire.} In the mood questionnaire, participants were asked a single question about their current well-being. The answers were mapped to a scalar mood score that was recorded for each answer. The score was used to track changes of participants' mood over time.

\item[\raisebox{-.15\height}{\includegraphics[height=8pt]{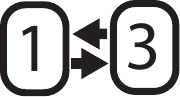}}]\textbf{Symbol Matching.} In the symbol matching test, participants were presented with a mapping of symbols to numbers. Participants were then prompted with a single symbol from this mapping, and asked to translate the shown symbol into the corresponding number using an on-screen virtual keyboard. Once the user entered their response, a new symbol would be shown. The goal was to translate the presented symbols as quickly and accurately as possible in a fixed amount of time. As metrics, the average response time and the number of correct responses were recorded. There was also a baseline version of this test in which participants simply had to input the presented numbers directly without any intermediate mapping for which the same metrics were recorded.

\item[\raisebox{-.15\height}{\includegraphics[height=10pt]{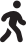}}]\textbf{Walking.} In the walking test, participants were asked to take a walk for two minutes, where ever they saw fit. Their smartphones recorded the number of steps taken during this walk in order to capture whether the participants' ability to walk was impaired.

\item[\raisebox{-.15\height}{\includegraphics[height=10pt]{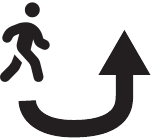}}]\textbf{U-turn.} In the U-turn test, participants were asked to walk as quickly as possible between two fixed points of their choice that should approximately have been four meters apart. Their smartphones recorded the number of turns and the average turn speed during this test in order to assess the participants' ability to turn.

\item[\raisebox{-.15\height}{\includegraphics[height=10pt]{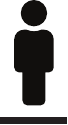}}]\textbf{Balance.} In the balance test, participants were asked to stand still and hold their balance for a fixed amount of time. The app recorded their postural sway during this test in order to evaluate to what degree the participant was able to remain still. Impaired postural control and an increased risk of falling are symptoms commonly associated with MS \cite{cameron2010postural}.

\item[\raisebox{-.15\height}{\includegraphics[height=10pt]{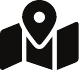}}]\textbf{Mobility.} The mobility test recorded the daily life space of the participant using their smartphone's location sensors. The mobility test was the only test that ran in the background and did not have to be manually activated.

\item[\raisebox{-.15\height}{\includegraphics[height=10pt]{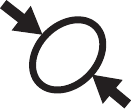}}]\textbf{Pinching.} In the pinching test, participants were presented with a series of virtual objects in varying locations on their smartphone screens. The participants were then asked to perform a pinching gesture using their fingers to squash the objects as quickly as possible. The app recorded the number of successfully squashed objects over a fixed amount of time, and which hand was used to perform the pinching gesture. The aim was to measure the participants' pinching ability which may be impaired in people with MS \cite{chen2007hand}.

\item[\raisebox{-.15\height}{\includegraphics[height=10pt]{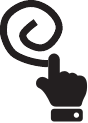}}]\textbf{Drawing.} In the drawing test, participants were asked to draw a sequence of shapes that was shown on their screens using their fingers - twice for each shape. The shapes represented a square, a circle, a figure eight, and a spiral. For each shape, the app recorded the best Hausdorff distance between the drawn and shown shapes.
\end{itemize}

\subsection{Problem Statement} We consider the setting in which we are given a number $k$ of tests, including test scores $s_i \in \mathbb{R}$, the time since the last test $t_i$ in seconds, a one-hot encoded representation $m_i \in \{0,1\}$ of the test metric type with $i \in [0 \isep k-1]$, and demographic data $d = (d_0. d_1)$, including age $d_0 \in \mathbb{N}_0$ and sex $d_1 \in \{0,1\}$, of the participant that performed the tests. Our goal is to train a predictive model $P$ that produces a scalar diagnostic score $y \in [0,1]$ that indicates the likelihood of the given set of test results belonging to a participant with or without MS.
\begin{align}
\label{eq:y}
y &= P([0, t_1, ..., t_i], [m_0, ..., m_i], [s_0, ..., s_i], d).
\end{align}
The primary challenge in this setting is to identify predictive patterns among the potentially large set of tests performed irregularly over long periods of time that provide evidence for or against a MS diagnosis.

\subsection{Attentive Aggregation Model (AAM)} As predictive model $P$, we use an AAM - a deep-learning model that utilises neural soft attention in order to integrate information from a potentially large number of smartphone test results. AAMs are based on evidence aggregation models (EAMs) \cite{schwab2019phonemd}. As a first step, we concatenate the time since the respective prior test $t_i$, result scores $s_i$, and test metric indicators $m_i$ from the smartphone tests into $k$ features $x_i$.
\begin{align}
x_i &= \text{concatenate}(t_i, m_i, s_i)
\end{align}
We then use a multilayer perceptron (MLP) with a configurable number $L$ of hidden layers and $N$ neurons per hidden layer to process each input feature into a $N$-dimensional high-level hidden feature representation $h_i$.
\begin{align}
h_i &= \text{MLP}(x_i)
\end{align}
Next, we aggregate the information from all $k$ high-level hidden representations into a single aggregated hidden representation $h_\text{all}$ that reflects all available tests for a given participant. To do so, we used a learned neural soft attention mechanism that weighs the individual hidden representations $h_i$ of each individual test instance by their respective importance $a_i$ towards the final diagnostic output score.
\begin{align}
h_\text{all} &= \textstyle{\sum}_{i=1}^{k} a_i h_i
\end{align}
Following \cite{schwab2018granger,schwab2017beat,xu2015show}, we calculate the attention factors $a_i$ by first projecting the individual hidden representations $h_i$ into an attention space representation $u_i$ through a single-layer MLP with a weight matrix $W$ and bias $b$.
\begin{align}
\label{eq:u_i}
u_{i} &= \text{activation}(Wh_{i} + b)
\end{align}
We finalise the calculation of the attention factors $a_i$ by computing a softmax over the attention space representations $u_i$ of $h_i$ using the most informative hidden representation $u_\text{max}$. $W$, $b$ and $u_{\text{max}}$ are learned parameters and optimised together with the other parameters during training \cite{schwab2019phonemd}. 
\begin{align}
\label{eq:a_i}
a_{i} &= \text{softmax}(u_{i}^Tu_{\text{max}})
\end{align}
We then calculate the final diagnostic score $y$ using a MLP with one sigmoid output node on a concatenation of the aggregated hidden state $h_\text{all}$ and the demographic data $d$. The integration of demographic information to the top-level MLP enables the model to adjust the final prediction $y$ for epidemiological risk factors \cite{pugliatti2006epidemiology}, such as sex and age.
\begin{align}
y &= \text{MLP}(\text{concatenate}(h_\text{all}, d))
\end{align}
AAMs rely solely on attention to aggregate test results over time. Using attention to perform the temporal aggregation has the advantage that attention mechanisms learn global input-output relations without regard to their distance in the input sequence, and in this manner improve learning of long-range temporal dependencies \cite{vaswani2017attention}. While global attention models are commonly employed in natural language processing  \cite{vaswani2017attention,devlin2018bert}, we are not aware of any prior works that apply global attention to improve learning of long-term temporal dependencies on smartphone sensor data.

\section{Experiments}
To evaluate the predictive performance of AAMs in diagnosing MS from smartphone data, we performed experiments that compared the diagnostic performance of AAMs and several baseline models on real-world smartphone monitoring data. Our experiments aimed to answer the following questions:
\begin{enumerate}
\item[1] What is the predictive performance of AAMs in diagnosing MS from smartphone data?
\item[2] What is the predictive performance of AAMs compared to other methods, such as Mean Aggregation and the demographic baseline?
\item[3] Which test types were most informative for diagnosing MS, and to what degree?
\item[4] To what degree does including more tests performed by subjects improve predictive performance of AAMs?
\item[5] Does the neural attention mechanism identify meaningful patterns?
\end{enumerate}
 
\subsection{Dataset and Study Cohort} We used data from the Floodlight Open study that recruited participants via their personal smartphones in, among others, the United States (US), Canada, Denmark, Spain, Italy, the Czech Republic, and Switzerland \cite{montalban2018floodlight,midaglia2019floodlight}. To perform our experimental comparison, we used all of the available smartphone monitoring data from April 23$^\text{rd}$ 2018 to August 29$^\text{th}$ 2019. In addition to regularly performing the smartphone-based tests, participants also provided their demographic profiles upon sign-up. The demographic profile included age, sex, and whether or not they had an existing diagnosis of MS. To ensure that a minimal amount of data points are available for diagnosis, we excluded all participants that had produced fewer than 20 test results during the analysed time frame (312 participants). We chose the cutoff at a minimum of 20 tests as this corresponds roughly to two sets of the daily test suite, which would be the minimal amount of data needed to assess symptom progression over time. We assigned the included patients to three folds for training (70\%), validation (10\%) and testing (20\%) randomised within strata of diagnostic status, app usage, sex, age, and number of tasks performed (Table \ref{tb:dataset}). The training, validation and test folds were respectively solely used for training, hyperparameter optimisation, and model evaluation. On average, participants with MS performed more tests over their app usage duration than participants without MS - presumably because of their intrinsic motivation to better understand their disease and help advance biomedical research.

\begin{table}[t!]
\caption{Population Statistics. \textup{Training, validation, and test fold statistics. Age and Usage are medians (10\% and 90\% quantiles in parentheses).}}
	\label{tb:dataset}
	\centering
	\begin{small}
		\begin{tabular}{l@{\hskip 0.45ex}l@{\hskip 0.5ex}l@{\hskip 0.5ex}l@{\hskip 0.5ex}r}
			\toprule
			Property & Training & Validation & Test \\
			\midrule
			Subjects (\#) & 542 (70\%) & 77 (10\%) & 155 (20\%)\\
			MS (\%) & 51.9 & 52.0 & 51.6\\
			Female (\%) & 60.3 & 59.7 & 60.7\\
			Age (years) & 41.0 (27.0, 59.0) & 41.0 (26.5, 56.5) & 41.0 (28.0, 57.5)\\
			Usage (days) & 22.4 (0.6, 203.5) & 19.0 (0.8, 175.5) & 18.8 (1.0, 130.0) \\
			\bottomrule
		\end{tabular}
	\end{small}
\end{table}

\subsection{Models} We trained a demographic baseline model, a Mean Aggregation baseline, and several ablations of AAMs. The AAMs used a fully-connected neural network as their base, and a single neuron with a sigmoid activation function as output. We trained one AAM version that received the demographic information (AAM + age + sex), and one that did not (AAM). For computational reasons, we limited the maximum number of test results per participant. To estimate the performance benefit of having access to information from more tests in the analysis, we trained AAMs and Mean Aggregation using up to the first 25, 30, 40, 50, 100, 150, 200, 250, 300 and 350 test results per participant, if available. For the demographic baseline, we used a random forest (RF) model that received as input only the age and sex of the participant (Age + sex). We used the demographic baseline to evaluate whether and to what degree data from the smartphone-based tests improves predictive performance, since the demographic baseline only had access to demographic data and did not include data from the smartphone-based tests. As a simple reference baseline, the Mean Aggregation utilised the mean normalised test result score to produce the final diagnostic score - this served to determine whether the use of more complex, learned aggregation methods, such as AAMs, is  effective and warranted. 

\subsection{Hyperparameters} To ensure all models were given the same degree of hyperparameter optimisation, we used a standardised approach where each model was given an optimisation budget of 50 hyperparameter optimisation runs with hyperparameters chosen from pre-defined ranges (Table \ref{tb:hyperparameters}). The best performing configuration on the validation fold was selected for further evaluation. For the demographic baseline model, we used a RF with $T$ trees and a maximum tree depth of $D$. For the AAMs, we used an initial MLP with $L$ hidden layers, $N$ hidden units per hidden layer, a dropout percentage of $p$ between hidden layers, and an L$2$ weight penalty of strength $s$. We trained AAMs to optimise binary cross entropy for a maximum of 300 epochs with a minibatch size of $B$ participants and a learning rate of 0.003. In addition, we used early stopping with a patience of 32 epochs on the validation fold. 

\begin{table}[b!] 
\caption{Hyperparameters. \textup{Ranges used for hyperparameter optimisation of AAMs (top) and the Age + sex baseline using a Random Forest model (RF, bottom). Parentheses indicate continuous ranges within the indicated limits sampled at uniform probability. Comma-delimited lists indicate discrete choices with equal selection probability.}}
	\label{tb:hyperparameters}
	\centering
	\begin{small}
		\begin{tabular}{l@{\hskip 2.5ex}l@{\hskip 11.5ex}r}
			\toprule
			& Hyperparameter & Range / Choices\\
			\midrule
			\parbox[t]{2mm}{\multirow{5}{*}{\rotatebox[origin=c]{90}{AAM}}}& Number of hidden units $N$ & 16, 32, 64, 128\\
			& Batch size $B$ & 16, 32, 64\\
			& L$2$ regularisation strength $s$ & 0.0001, 0.00001, 0.0\\
			& Number of layers $L$ & (1, 3)\\
			& Dropout percentage $p$ & (0\%, 35\%)\\
			\midrule
			\parbox[t]{2mm}{\multirow{2}{*}{\rotatebox[origin=c]{90}{RF}}}&Tree depth $D$ & 3, 4, 5\\
			& Number of trees $T$ & 32, 64, 128, 256\\
			\bottomrule
		\end{tabular}
	\end{small}
\end{table}

\begin{table*}[t!] 
\caption{Predictive Performance. \textup{Comparison of Attentive Aggregation Models (AAMs), Mean Aggregation, and a demographic baseline (Age + sex) in terms of AUC, AUPR, F$_1$, sensitivity, and specificity for predicting MS on the test set using a maximum of 250 test results from each participant. In parentheses are the 95\% confidence intervals (CIs) obtained via bootstrap resampling. \sig = significant at p < 0.05 to AAM + age + sex.}}
	\label{tb:results_all}
	\centering
	\begin{small}
		\begin{tabular}{l@{\hskip 2.5ex}l@{\hskip 3ex}l@{\hskip 3ex}l@{\hskip 3ex}l@{\hskip 3ex}l}
			\toprule
			Model (max. 250 test results) & AUC & AUPR & F$_1$ & Sensitivity & Specificity\\
			\midrule
			AAM + age + sex & \hspace{1ex}\textbf{0.88} (0.70, 0.88) & \hspace{1ex}\textbf{0.90} (0.67, 0.90) & \hspace{1ex}\textbf{0.80} (0.65, 0.83) & \hspace{1ex}\textbf{0.83} (0.59, 0.86) & \hspace{1ex}{0.73} (0.62, 0.89) \\ 
			Mean Aggregation + age + sex  & \sig0.77 (0.70, 0.82) & \sig0.76 (0.64, 0.84) & \sig0.71 (0.65, 0.78) & \sig0.68 (0.59, 0.85) & \sig{0.75} (0.59, 0.87) \\ 
			Age + sex & \sig{0.76} (0.69, 0.84) & \sig{0.75} (0.65, 0.86) & \sig{0.69} (0.62, 0.79) & \sig{0.73} (0.55, 0.83) & \sig{0.61} (0.58, 0.89) \\
			AAM & \sig{0.72} (0.56, 0.82) & \sig{0.67} (0.57, 0.84) & \sig{0.61} (0.53, 0.77) & \sig{0.63} (0.45, 0.79) & \sig{0.83} (0.53, 0.86) \\ 
			Mean Aggregation & \sig0.56 (0.50, 0.67) & \sig0.61 (0.49, 0.74) & \sig0.39 (0.20, 0.54) & \sig0.28 (0.13, 0.43) & \sig\textbf{0.85} (0.70, 0.97) \\ 
			\bottomrule
		\end{tabular}
	\end{small}
\end{table*}

\subsection{Preprocessing} We normalised the time between two test results $t_i$ to the range $[0,1]$ using the highest observed $t_i$  on the training fold of the dataset. We additionally normalised all test result scores $x_i$ to the range of $[0,1]$ using the lowest and highest observed test result for each test metric on the training fold of the dataset.

\subsection{Metrics}
\subsubsection{Predictive Performance} 
We evaluated all models in terms of their area under the receiver operating characteristic curve (AUC), the area under the precision recall curve (AUPR), and F$_1$ score on the test fold of 155 participants. For the comparison of predictive performance, we additionally computed the sensitivity and specificity of the respective models. We also quantified the uncertainty of all the performance metrics by computing 95\% confidence intervals (CIs) using bootstrap resampling with 1000 bootstrap samples. To assess the statistical significance of our results, we applied Mann-Whitney-Wilcoxon (MWW) tests at significance level $\alpha = 0.05$ to the main comparisons. We additionally applied the Bonferroni correction to adjust for multiple comparisons.

\begin{figure}[b]
	\centering
	\includegraphics[width=0.9\linewidth]{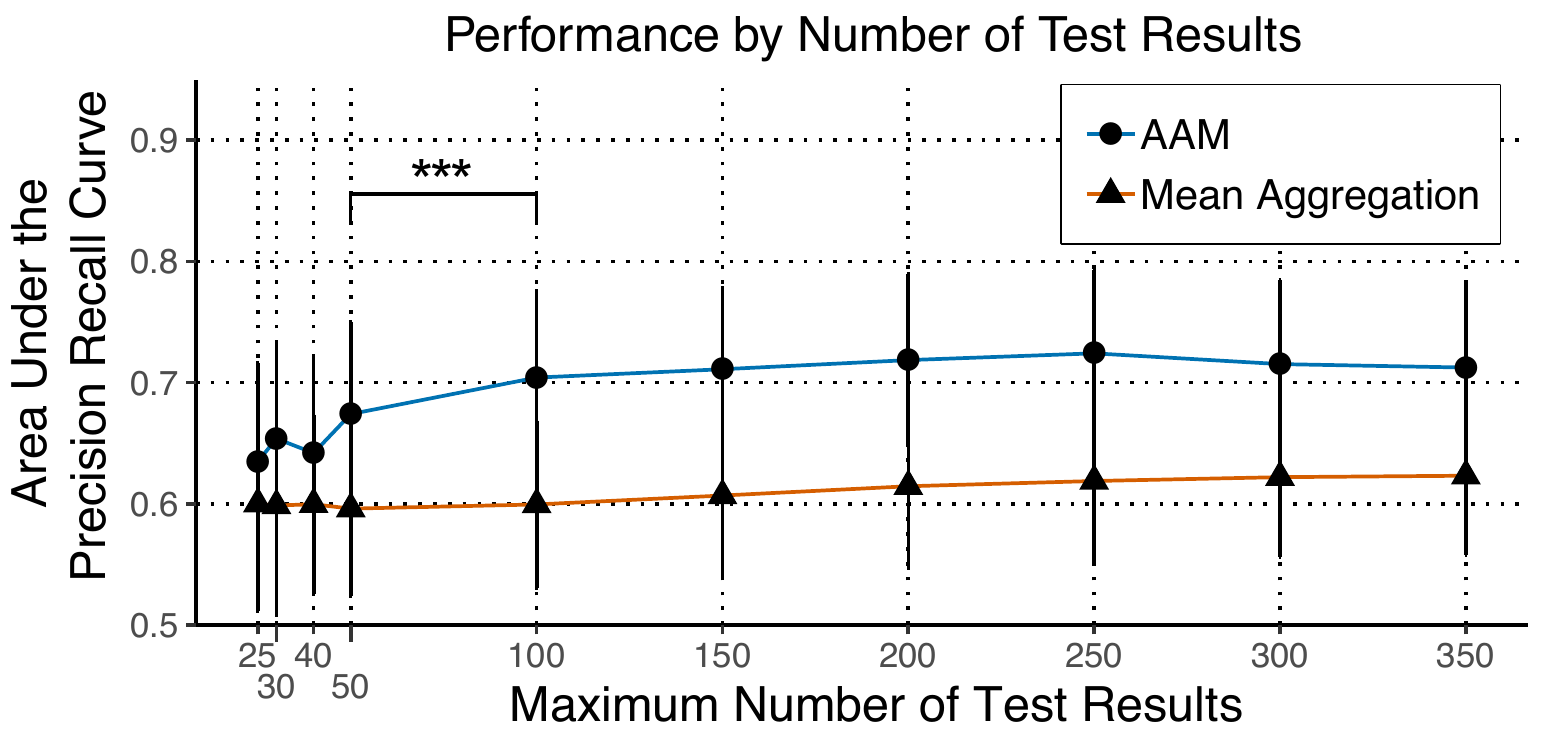}
	\caption{Performance comparison of AAM (dots, blue) and Mean Aggregation (triangles, orange) in terms of their Area Under the Precision-Recall Curve (AUPR, y-axis) when varying the maximum number of test results (x-axis) available to predict the MS diagnosis for each participant. *** = significant at p < 0.001.}
	\label{fig:task_number}
\end{figure}
\subsubsection{Importance of Test Types}
To quantify the importance of the various test types toward the diagnostic performance of the AAM, we retrained AAMs with the same hyperparameters after removing the test results from exactly one type of test. The reduction in predictive performance associated with removing the information of one test type can be seen as a proxy for the importance of that test type \cite{schwab2018granger,schwab2019cxplain}, since features that are associated with a higher reduction in prediction error carry more weight in improving the model's ability to predict MS.
\subsubsection{Neural Attention}
In order to qualitatively inspect the patterns that were captured by the AAM in the data, we additionally plotted the attention assigned to the test results from a sample participant with MS over time.

\begin{figure}[b]
	\centering
	\includegraphics[width=0.9\linewidth]{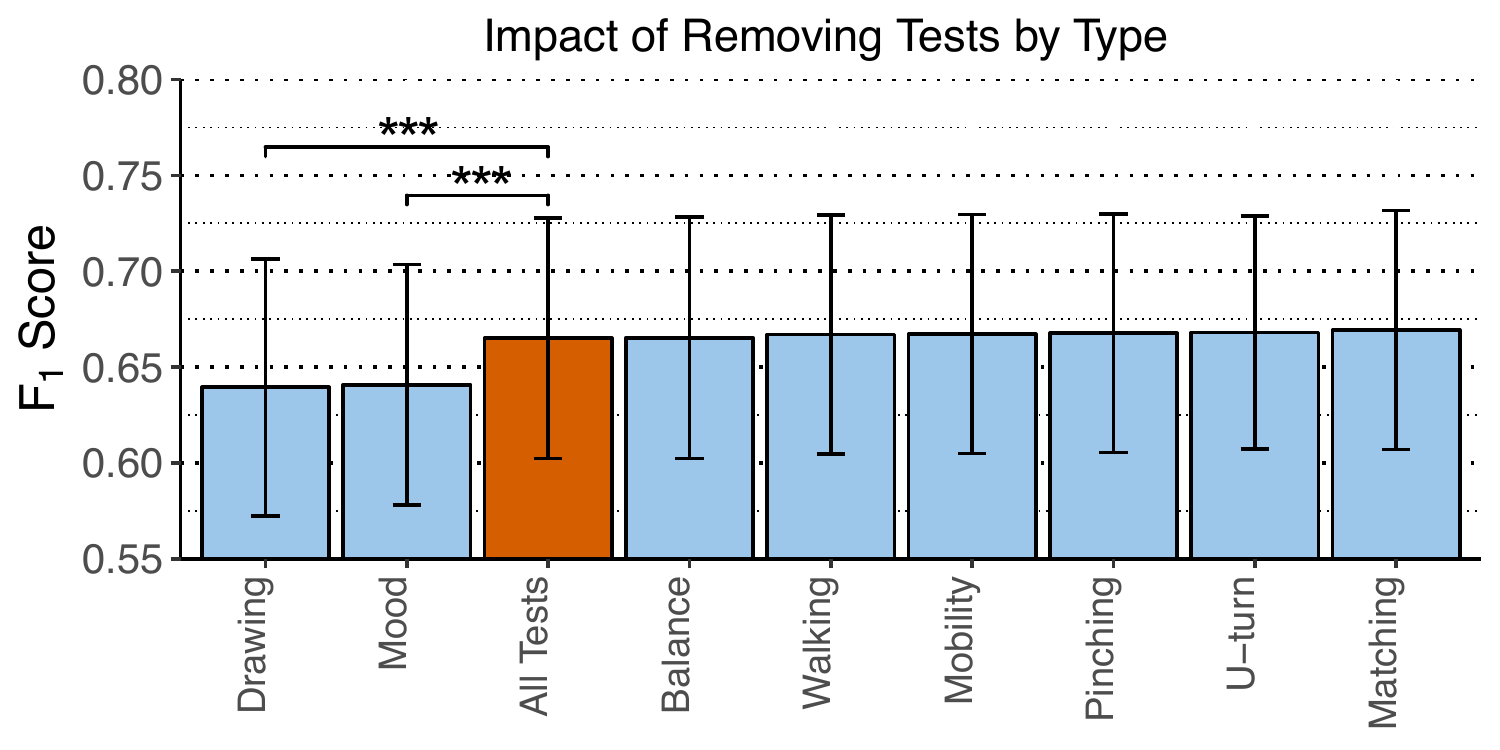}
	\caption{Performance comparison of AAM in terms of their F$_1$ score (y-axis) in predicting the MS diagnosis for each participant after removal of the information of all tests of a specific type (labelled test types, bottom) from the dataset. The reference baseline without removal of any test types (All Tests) is highlighted in orange. *** = significant at p < 0.001.}
	\label{fig:test_performance}
\end{figure}

\begin{figure*}[t]
	\centering
	\includegraphics[width=0.775\linewidth]{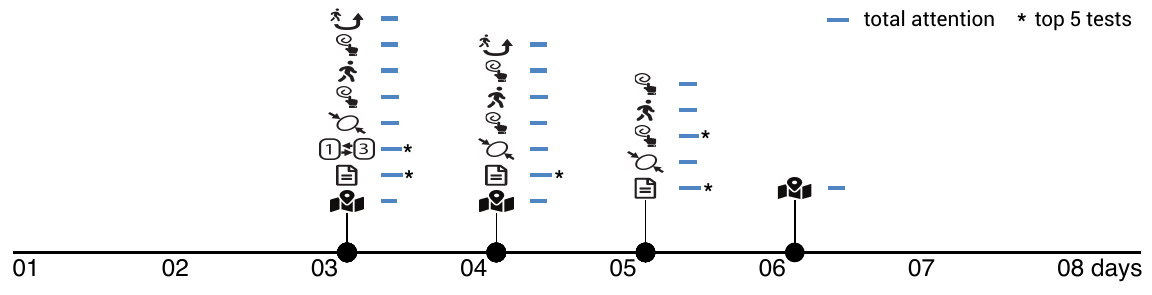}
	\caption{A set of test results as performed by a female participant (FL51683656) aged 49 with MS from the test set. The timeline depicts usage days from earlier (left) to later (right). The black dots indicate that at least one test was performed that day. The absence of a dot indicates that no test was performed that day. Symbols connected to marked days depict which tests were performed (Figure \ref{fig:teaser}). The blue bars directly adjacent to the symbols show how much total attention was assigned to that test instance. Total attention refers to the sum of attention assigned to all test results  belonging to a test instance, e.g. the Hausdorff distances for all shapes in a drawing test. We found that the model focused predominantly on the mood, symbol matching, and drawing tests (* marks top 5 tests) to correctly identify (score = 0.77, threshold = 0.49) that this participant has MS.}
	\label{fig:atn}
\end{figure*}

\section{Results}
\subsection{Predictive Performance} 
In terms of predictive performance for diagnosing MS, the AAMs models with demographic information (AAM + age + sex) achieved a significantly ($p < 0.05$) higher AUC, AUPR and F$_1$ than all the baselines we evaluated (Table \ref{tb:results_all}). In particular, we found that integrating the information from the smartphone tests was crucial, as AAMs that used both the demographic data and the smartphone test information significantly ($p < 0.05$) outperformed the demographic baseline (Age + sex) in terms of AUC, AUPR, and F$_1$. Mean Aggregation had a considerably lower performance than AAMs, and displayed the worst AUC, AUPR and F$_1$ of the compared models - demonstrating that the use of more sophisticated adaptive aggregation models for integrating information from smartphone-based tests over time, such as AAMs, is effective and warranted. We also found that AAMs achieved a high level of both sensitivity and specificity, whereas the demographic model emphasised sensitivity, and Mean Aggregation specificity. As expected, the AAM achieved a significantly higher performance when it had access to demographic information, as it was able to adjust for demographic risk factors by, for example, assigning higher risk to female app users - which have an up to three times higher MS risk than men \cite{pugliatti2006epidemiology}.

\subsection{Impact of Performing More Tests}

The results from comparing AAMs and Mean Aggregation in terms of their AUPR across a wide range of numbers of test results indicated that AAMs are better able to leverage an increasing number of tests performed (Figure \ref{fig:task_number}). In terms of predictive performance as measured by the AUPR, the AAM surpassed the Mean Aggregation baseline at all evaluated maximum numbers of test results. We also found that having access to a higher number of tests consistently and significantly (50 vs. 100 test results, $p<0.001$) improved the performance of AAMs up to a maximum of 250 test results per participant. Our results indicate that a respectable level of predictive performance could be achieved after collecting just 100 test results. In practice, data collection could for example be initiated upon referral to a specialist clinic or between the first clinical contact and the diagnostic appointment. We note that the time required to collect 100 test results is small in relation to the median time between symptom onset and diagnosis, which may be over two years in some countries \cite{fernandez2010characteristics}. In addition, Floodlight Open tests have been reported to be well accepted by patients with good adherence and patient satisfaction over a monitoring period of 24 weeks \cite{midaglia2019floodlight}.

\subsection{Importance of Test Types} 
When comparing the marginal reduction in prediction error associated with removing a specific type of test from the set of available tests, we found that the drawing and mood tests were contributing significantly larger marginal reductions ($p<0.001$) in prediction error to the AAM (Figure \ref{fig:test_performance}) - indicating that the drawing and mood tests were more predictive of MS diagnosis than other test types. The removal of other test types did not lead to similarly considerable reductions in prediction error compared to the AAM that had access to information from all test types (All Tests). This result could indicate that the results of other test types were either (i) not highly predictive of MS, or (ii) correlated with other tests to a degree that strongly impacted their marginal contributions.  Further prospective studies will be necessary to determine the optimal set of tests because confounding factors, such as different usage patterns in response to different test suites being available, may also influence predictive performance.

\subsection{Neural Attention}
Qualitatively, on the data from one sample participant, we found that the model focused on mood, drawing, and symbol matching tests to diagnose MS (Figure \ref{fig:atn}). The focus on mood and drawing tests for this subject are in line with our findings on the overall importance of the test types (Table \ref{fig:test_performance}).

\section{Discussion}
To the best of our knowledge, this work is the first to present a machine-learning approach to diagnosing MS from long-term smartphone monitoring data collected outside of the clinical environment. To derive a scalar diagnostic score for MS from smartphone monitoring data, we used an AAM to aggregate data from multiple types of tests over long periods of time. The AAMs used neural attention to quantify the degree to which individual tests contribute to the final diagnostic score, and to overcome the challenges of missingness, sparse data, long-term temporal dependencies and irregular sampling. Our experimental results indicate that AAMs outperform several strong baselines, identify meaningful data points within the set of performed tests, and that smartphone-based digital biomarkers could potentially be used to aid in diagnosing MS alongside existing clinical tests. Among the several potential advantages of using smartphone-derived digital biomarkers for diagnosing MS are that smartphone-based tests (i) can be administered remotely and therefore potentially expand access to underserved geographic regions, (ii) are inexpensive to distribute and could therefore potentially become a low-cost alternative to more expensive in-clinic tests, and (iii) are able to integrate information from long-term symptom monitoring, and therefore potentially better represent and quantify fluctuations in symptom burden over time. An additional benefit of using smartphone-based diagnostics for MS is that machine-learning models can, as demonstrated in this work, identify which symptom categories are most indicative of MS and how they interact over time, and could therefore potentially be used to monitor disease progression and inform follow-up treatment decisions. Finally, our results show that information from smartphone-based biomarkers are to some degree orthogonal to more traditional measurements, such as demographic data, and could therefore potentially be integrated with information from existing clinical tests and other multimodal data sources, such as MRI, to further increase diagnostic accuracy.

\begin{figure}[t]
	\centering
	\includegraphics[width=0.773\linewidth]{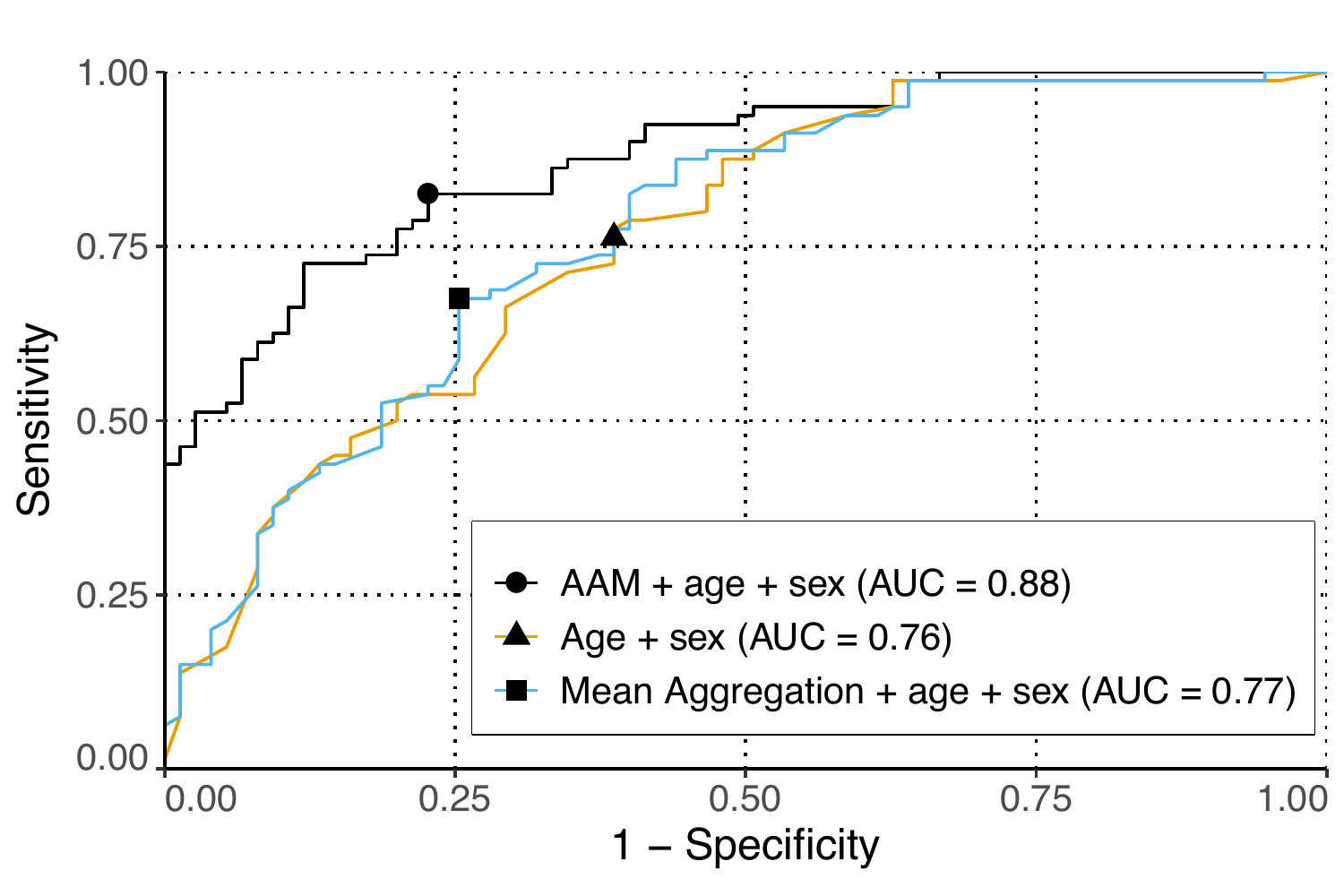}
	\caption{Comparison of the receiver operating characteristic (ROC) curves of AAM + age + sex (black, dot), Age + sex (orange, triangle), and Mean Aggregation + age + sex (blue, square) when using a maximum of 250 test results to predict the MS diagnosis for each test set participant. Symbols indicate the operating points (thresholds selected on the validation fold) presented in the comparison in Table \ref{tb:results_all}. }
	\label{fig:roc}
\end{figure}

\subsection{Limitations} While of respectable size, the studied cohort was restricted to residents of a limited number of countries and likely not representative of the global population. More importantly, the data originated from patients with a diagnosis at various stages of the disease, instead of pre-diagnosis. Therefore, our work cannot conclude whether people without clinical MS diagnosis could be identified before they receive their diagnosis. A prospective validation in a larger, clinically representative cohort will be necessary to conclusively establish the performance and utility of smartphone-derived biomarkers as a tool to aid in the diagnosis of MS, and their robustness when confronted with other disorders that have similar symptoms. Further work should investigate whether such biomarkers are also suitable to track disease progression, identify and predict relapses, enable the effective tuning of therapeutic options and medication dosages, enable earlier diagnoses, and eventually also provide an accurate prognosis. 

\section{Conclusion}
We presented a novel machine-learning approach to distinguishing between people with and without MS based on long-term smartphone monitoring data. Our method uses an AAM to aggregate data from multiple types of tests over long periods of time to produce a scalar diagnostic score. AAMs use neural attention to quantify the degree to which individual tests contributed to the diagnostic score, and to overcome the challenges of missingness, sparse data, long-term temporal dependencies and irregular sampling. In an experimental evaluation on smartphone monitoring data from a cohort of 774 people, we demonstrated that AAMs identify predictive and meaningful digital biomarkers for diagnosing MS. Our experiments show that smartphone-derived digital biomarkers could potentially be used to aid in diagnosing MS in the future alongside existing clinical tests. Smartphone-based tools for tracking symptoms and symptom progression in MS may improve clinical decision-making by giving clinicians access to objectively measured high-resolution health data collected outside the restricted clinical environment. In addition, our solution based on attention mechanisms further elucidates the basis of the model decisions and may enhance the clinician's understanding of the provided diagnostic score. We therefore believe our initial results may warrant further research on how digital biomarkers could be integrated into clinical workflows.

\appendices

\section*{Acknowledgments}
The data used in this manuscript were contributed by users of the Floodlight Open mobile application developed by Genentech Inc: \url{https://floodlightopen.com}. Patrick Schwab is an affiliated PhD fellow at the Max Planck ETH Center for Learning Systems.

\small
\bibliographystyle{IEEEtran}
\bibliography{references.bib}

\end{document}